\documentclass[a4paper,abstract=on, enabledeprecatedfontcommands]{scrartcl}

\usepackage{geometry}
\usepackage{amssymb, amsmath, amsthm}
\usepackage[pdftex]{graphicx}
\usepackage{enumerate}

\title{\textsc{Classification of chemical compounds based on the correlation between \textit{in vitro} gene expression profiles}\thanks{This study is supported by JSPS KAKENSHI Grant Numbers JP20K12203.}}

\author{Jun-ichi Takeshita%
  \thanks{Research Institute of Science for Safety and Sustainability,
    National Institute of Advanced Industrial Science and Technology (AIST),
    Tsukuba,
    Japan. (\texttt{jun-takeshita@aist.go.jp})}
  \and
  Akinobu Toyoda%
  \thanks{School of Science and Engineering,
    University of Tsukuba, Tsukuba, Japan.}
  \and
  Hidenori Tani%
  \thanks{Environmental Management Research Institute, %
    National Institute of Advanced Industrial Science and Technology (AIST), 
    Tsukuba, Japan. }
  \and
  Yasunori Endo
  \thanks{Faculty of Systems and Information Engineering,
    University of Tsukuba, Tsukuba, Japan.}
  \and
  Sadaaki Miyamoto
  \thanks{Faculty of Systems and Information Engineering,
    University of Tsukuba, Tsukuba, Japan.}
}

\begin{document}
\maketitle

\begin{abstract}
  Toxicity evaluation of chemical compounds has traditionally relied on animal experiments;
however, the demand for non-animal-based prediction methods for toxicology of compounds is increasing worldwide.
Our aim was to provide a classification method for compounds based on \textit{in vitro} gene expression profiles.
The \textit{in vitro} gene expression data analyzed in the present study was obtained from our previous study.
The data concerned nine compounds typically employed in chemical management.
We used agglomerative hierarchical clustering to classify the compounds;
however, there was a statistical difficulty to be overcome.
We needed to properly extract RNAs for clustering from more than 30,000 RNAs.
In order to overcome this difficulty, we introduced a combinatorial optimization problem with respect to both gene expression levels and the correlation between gene expression profiles.
Then, the simulated annealing algorithm was used to obtain a good solution for the problem.
As a result, the nine compounds were divided into two groups using 1,000 extracted RNAs.
Our proposed methodology enables read-across, one of the frameworks for predicting toxicology, based on \textit{in vitro} gene expression profiles.

\noindent
\textbf{keywords}:
Statistical classification, 
Multiobjective combinatorial optimization,
Chemical toxicity,
Alternatives to animal experiments,
\textit{In vitro} gene expression,
Mathematical formulation
\end{abstract}

\section{Introduction}
\label{intro}
Traditionally, toxicity evaluation of chemical compounds has relied on animal experiments~[\cite{Eaton2015}].
However, in terms of time, cost efficiency, and animal welfare concerns, there is an increasing demand for the development of non-animal-based methodologies for predicting chemical toxicity.
Recently, several elements have been proposed for use in toxicity prediction methods, such as (quantitative) structure-activity relationships ((Q)SAR), quantitative activity-activity relationships (QAAR), and read-across~[\cite{OECD2014}].

Read-across is a method whereby the toxicity of a given compound is predicted without the use of animal test data, but instead based on the animal toxicity data for similar compounds.
One study, for example, conducted read-across based on similarity in chemical structure and toxicology expert judgment~[\cite{Mellor2017}].
For read-across, it is important to properly group compounds using data that does not rely on animal experiments.
In order to facilitate the grouping of compounds, the Organisation for Economic Co-operation and Development (OECD) published the OECD QSAR Toolbox~[\cite[Section 4.3]{Madden2013}], [\cite{OECD2017}].
In addition, the Japanese government and participating academic institutes developed the Hazard Evaluation Support System Integrated Platform (HESS)~[\cite{Sakuratani2013}], [\cite[Section 4.4]{Madden2013}], [\cite{NITE2017}].
These platforms group compounds based on \textit{in silico} parameters, that is, chemical structures and essential physicochemical parameters;
and by employing them, we can successfully group compounds based on these parameters.

In order to predict toxicity, it is useful to use not only \textit{in silico} parameters but also \textit{in vitro} parameters, because the latter can reflect certain biological characteristics of compounds.
In fact, in the case of predicting hepatotoxicity, one of the most prevalent forms of toxicity, two studies have reported that using \textit{in vitro} parameters increased the accuracy of discriminative models for predicting the presence or absence of hepatotoxicity, compared to using \textit{in silico} parameters~[\cite{Liu2015}], [\cite{Low2011}].
These studies suggest the hypothesis that grouping compounds based on \textit{in vitro} parameters would increase the accuracy of read-across approaches for predicting chemical toxicity, compared to grouping based on \textit{in silico} parameters.

The present study proposes a methodology for grouping compounds using \textit{in vitro} gene expression data.
In order to group compounds, agglomerative hierarchical clustering was applied;
however, a statistical difficulty appeared.
The gene expression data, from which it was necessary to properly extract a limited number of RNAs for clustering compounds, included more than 30,000 RNAs.
In order to overcome this difficulty, we introduced a multiobjective combinatorial optimization problem with respect to both gene expression levels and the correlation between gene expression profiles.
Then, we applied the simulated annealing algorithm, a metaheuristic algorithm, to obtain a good solution for the multiobjective combinatorial optimization problem.

\section{Methods and materials}
\label{sec:1}

\subsection{Gene expression data}
The present study used the data reported in~\cite{Tani}.
In that study, cell-based assays were conducted, in duplicate, using mouse embryonic stem cells, for nine compounds typically employed in chemical management: bis-phthalate, $p$-dicholorobenzene, phenol, trichloroethylene, benzene, chloroform, $p$-cresol, and tri-$n$-butyl-phosphate.
In other words, each compound had two samples, and each group of nine samples had the same control condition.
Then, the gene expression levels were quantified using the fragments per kilobase of exon per million mapped fragments (FPKMs), and reported for a total of 32,586 RNAs.
We used these RNAs and their FPKMs in the present study's analysis.

\subsection{Gene expression ratio}
For any compound-treated group, the following gene expression ratio was used for each gene:
\begin{equation*}
  f (x, y) = \log_2 \left(\frac{y}{x}\right),
\end{equation*}
where $x$ and $y$ denote the gene expression levels of the control and compound-treated groups, respectively.
Note that if $x$ or $y$ was zero, then the next smallest value in the respective group was used.

\subsection{Clustering}
In order to group compounds, agglomerative hierarchical clustering (the average linkage between the merged groups) was used, because this method can be used for any dissimilarity measures.
The following dissimilarity measure was used for the hierarchical clustering:
for any two compounds, $x$ and $y$, the dissimilarity measure, or distance, between $x$ and $y$, say, $d(x,y)$ was defined by
\begin{equation*}
  d(x,y) = \frac{1 - \text{corr} (x, y)}{2},
\end{equation*}
where $\mbox{corr} (x, y)$ is the correlation coefficient between the respective FPKM vectors of Compound $x$ and Compound $y$.
The dissimilarity measure takes a value between $0$ and $1$.

\subsection{Selection of RNAs}
In order to extract RNAs that clearly revealed the difference between compounds, the present study introduced the following combinatorial optimization problem to extract $n$ RNAs for a given natural number $n$:
\begin{align*}
  &\text{objective function}\\
  & \quad U = (1 - \alpha) U_1 + \alpha U_2  \ (0 \leq \alpha \leq 1),\\
  & \quad U_1 = \frac{1}{\text{Count} (w)} \times \\
  & \qquad \sum_{x, y \in E, x \neq y} w_{x, y}\frac{\left| (1/n) \sum_{i=1}^{n} (x_i - \overline{x}) (y_i - \overline{y}) \right|}%
  {\left\{ (1/n) \sum_{i=1}^{n} (x_i - \overline{x})^2\right\}^{1/2} \left\{ (1/n) \sum_{i=1}^{n} (y_i - \overline{y})^2\right\}^{1/2}},\\
  & \quad U_2 = \frac{1}{n} \sum_{i=1}^{n} \left(\frac{\|r_i\|}{\max_j \|r_j\|} \right),\\
  &\mbox{subject to\quad}  \{x_i\}, \{y_i\} \subset \Gamma \mbox{ and } \#\{x_i\} = \#\{y_i\} = n,
\end{align*}
where $E$ and $\Gamma$ denote the set of compound-treated groups and the set of all the RNAs (32,586 RNAs), respectively;
$x_i$ and $y_i$ denote the expression ratios of RNA${}_i$ for $x$ and $y \in E$, respectively;
$\overline{x}$ and $\overline{y}$ denote the means of $x_i$ and $y_i$, respectively;
$w_{x,y}$ denotes weights and takes a value in $\{-1, 0, 1\}$;
and $\text{Count}(w)$ denotes the number of $w_{x,y}$ taking a number of $1$.
Note that, for some $x$ and $y$, if $w_{x,y} = 1$, we can extract RNAs that increase the correlation between Compound $x$ and Compound $y$;
and if $w_{x,y} = -1$, we can extract RNAs that decrease the correlation between Compound $x$ and Compound $y$;
otherwise, we are not interested in the correlation between the two compounds.
In addition, $r_i$ designates the FPKM vector for RNA${}_i$, and $\|\cdot\|$ represents the Euclidean norm.

The function $U_1$ takes a value in $[0,1]$, since the sigma component is divided by $\text{Count}(w)$.
The function $U_2$ also takes a value in $[0,1]$.
Thus, the objective function $U$ takes a value in $[0,1]$, since $\alpha$ is a parameter taking a value in $[0,1]$.
Roughly speaking, the function $U_1$ describes the strength of the correlation between two compound-treated groups, $x$ and $y$, based on extracted RNAs.
The function $U_2$ describes the gene expression level of extracted RNAs.
Then, the function $U$ is a linear combination of $U_1$ and $U_2$.

In the present study, the simulated annealing algorithm, which was originally introduced by \cite{Kirkpatrick1983} and \cite{Cerny1985}, was used to obtain a good solution for the combinatorial optimization problem.
Let $n$ be the number of RNAs we want to extract, $T$ be the initial temperature, $T_t \ (0 < T_t < T)$ be the final temperature, and $\gamma \ (0 < \gamma <1)$ be the cooling rate.
Then, the following is the algorithm to extract RNAs.
\begin{description}
  \item[Step 1] Choose $n$ RNAs randomly as the initial state.
  Let $R$ be the set of the chosen $n$ RNAs, and then calculate the objective function $U$ using the set $R$.
  In addition, set the initial temperature $T$.
  \item[Step 2] As a neighbor solution, generate a set $R'$, in which random elements of $R$ and $\overline{R}$ (the complement set of $R$) are exchanged.
  Then, calculate the objective function $U'$ using the set $R'$.
  \item[Step 3] If $U < U'$, then $R:= R'$; otherwise $R:=R'$ with the following probability $P$:
  \begin{equation*}
    P = \exp\left(- \frac{|U' - U|}{T}\right).
  \end{equation*}
  \item[Step 4] $T := \gamma T.$
  If $T \geq T_t$, output the set $R$ as the final state; otherwise, go to Step 2.
\end{description}

\section{Results and discussion}
\subsection{Grouping compounds using all the RNAs}
\begin{figure}[htbp]
  \includegraphics[width=0.75\textwidth]{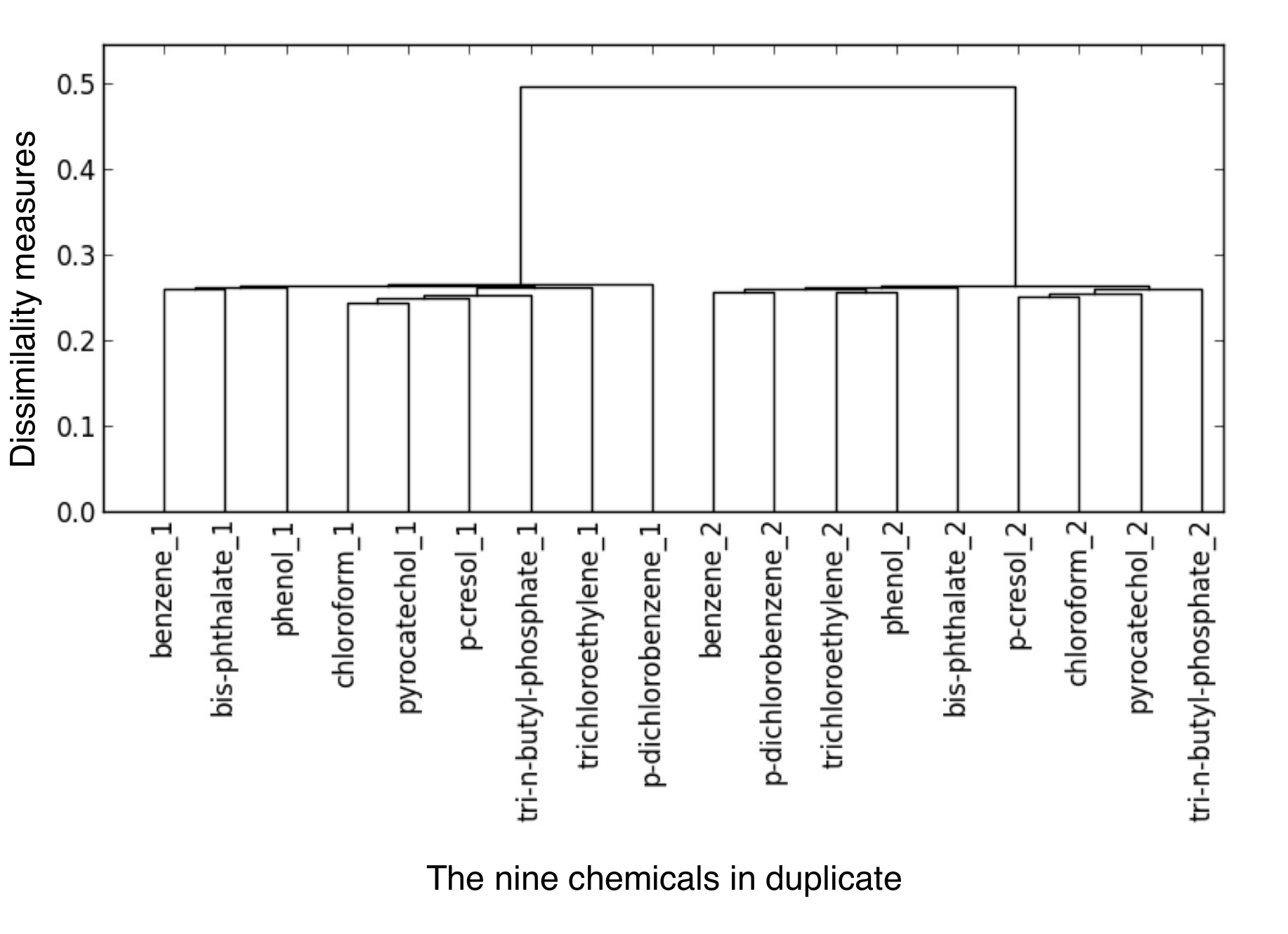}
\caption{A dendrogram obtained by applying aggregative hierarchical clustering (the average linkage between the merged groups) to the data of the gene expression ratios for the nine compounds and 32,586 RNAs.
The $y$-axis marks the dissimilarity measures at which the clusters merge, and the $x$-axis the distribution of the nine compounds in duplicate.}
\label{fig:1}       
\end{figure}

\begin{figure}[htbp]
  \includegraphics[width=0.75\textwidth]{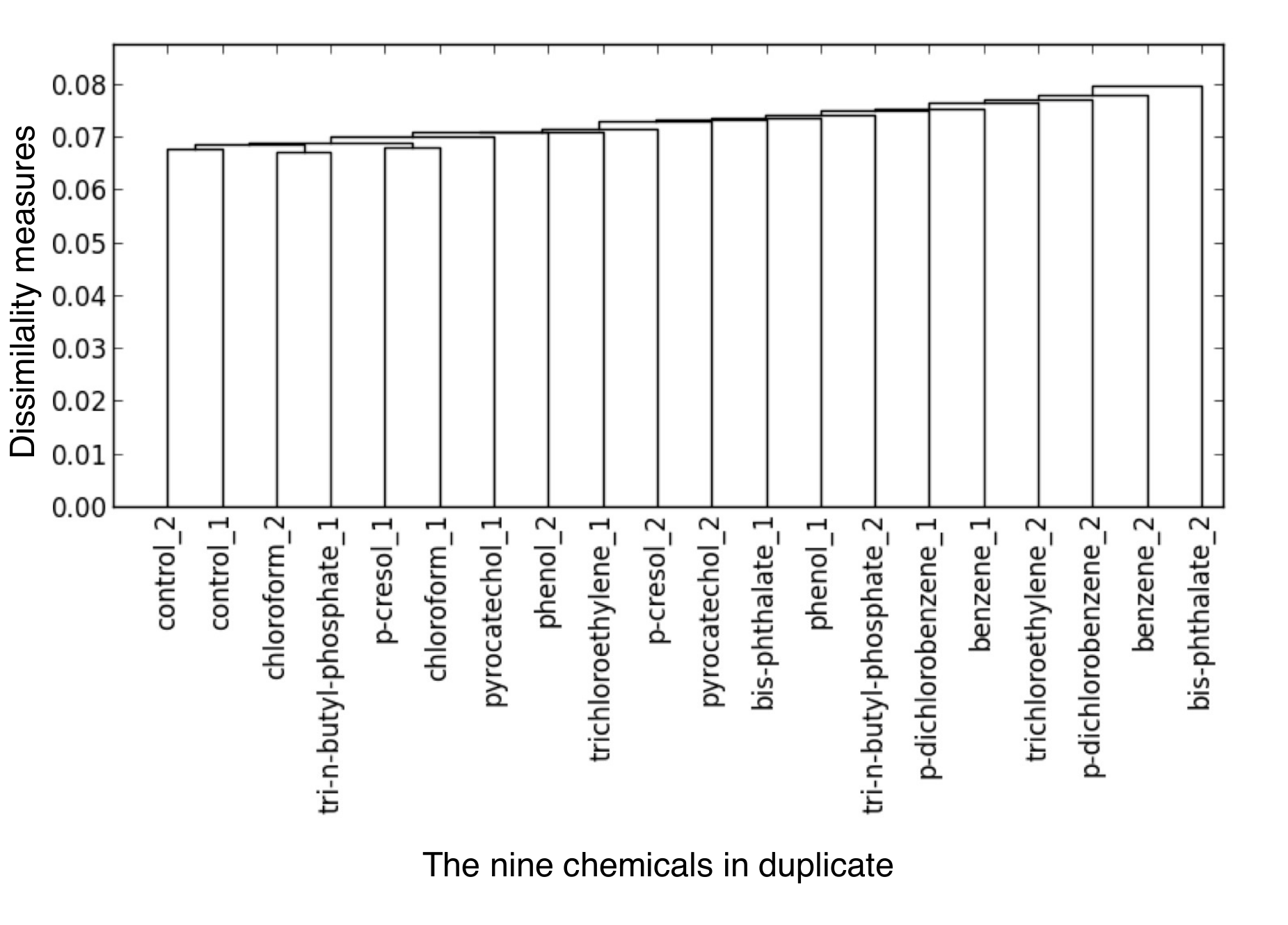}
\caption{A dendrogram obtained by applying aggregative hierarchical clustering (the average linkage between the merged groups) to the data of the gene expression levels for the nine compounds and 32,586 RNAs.
The $y$-axis marks the dissimilarity measures at which the clusters merge, and the $x$-axis the distribution of the nine compounds in duplicate.}
\label{fig:2}       
\end{figure}

Figure~\ref{fig:1} shows a dendrogram obtained by applying agglomerative hierarchical clustering (the average linkage between the merged groups) to the data set of gene expression ratios, and Figure~\ref{fig:2} a dendogram obtained by applying similar clustering to the set of gene expression levels, for the nine compounds in duplicate and all the RNAs (32,586 RNAs).
In both figures, the $y$-axes show the dissimilarity measures where two clusters were merged, while the $x$-axes show the distribution of compounds.
Note that subscripts 1 and 2 refer to the sample numbers; that is, if any two compounds have the same subscript number, the control conditions for the two compounds are identical.

In Figure~\ref{fig:1}, we can see that there are two robust clusters, and each cluster consists of the nine compounds with the same subscript number.
This implies that the two compound groups strongly depend on the gene expression levels in the control conditions.
In Figure~\ref{fig:2}, the two control conditions are closely proximate, but there is no significant difference between the 18 samples (the nine compounds in duplicate).
These results indicate that there is no significant difference between the gene expression patterns of the two control conditions.
Therefore, we may infer that the significant difference between the two control conditions in Figure~\ref{fig:1} was the result of the incremental accumulation of difference among the RNAs, as more than 30,000 RNAs were in the data set.
These results indicate that it is unreliable to use the information of all the RNAs for classifying compounds, but instead we must extract a limited number of RNAs in order to properly classify compounds using \textit{in vitro} gene expression data.

\subsection{Grouping compounds using extracted RNAs}
\begin{figure}[htbp]
  \includegraphics[width=\textwidth]{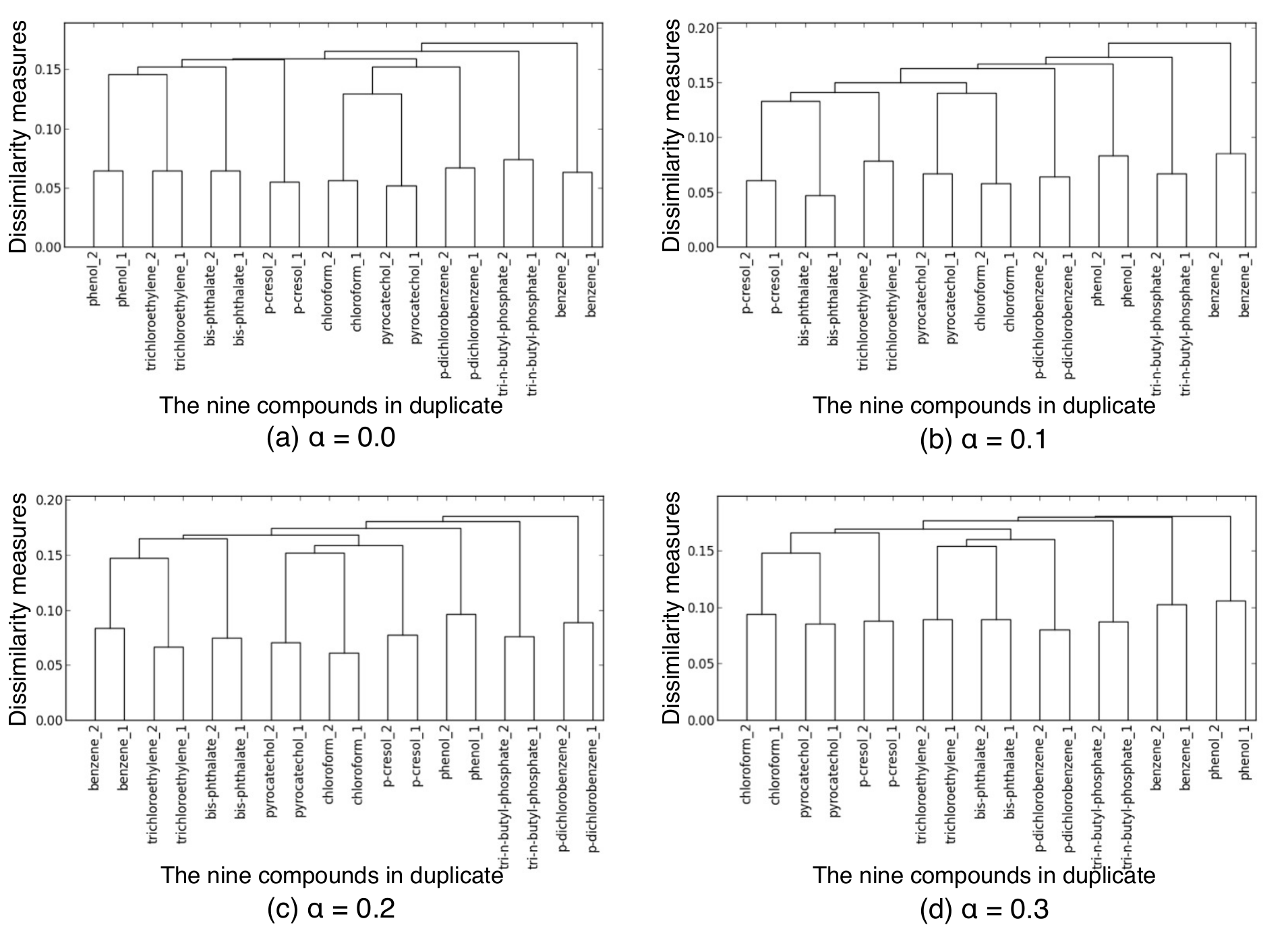}
\caption{Four dendrograms obtained by applying aggregative hierarchical clustering (the average linkage between the merged groups) to the data of the gene expression ratios for the nine compounds and $3,000$ extracted RNAs.
The upper-left (a), upper-right (b), lower-left (c) and lower-right (d) panels are the cases of $\alpha = 0.0, 0.1, 0.2$, and $0.3$, respectively.
In each panel, the $y$-axis marks the dissimilarity measures at which the clusters merge, and the $x$-axis the distribution of the nine compounds in duplicate.}
\label{fig:3}       
\end{figure}

\begin{figure}[htbp]
  \includegraphics[width=\textwidth]{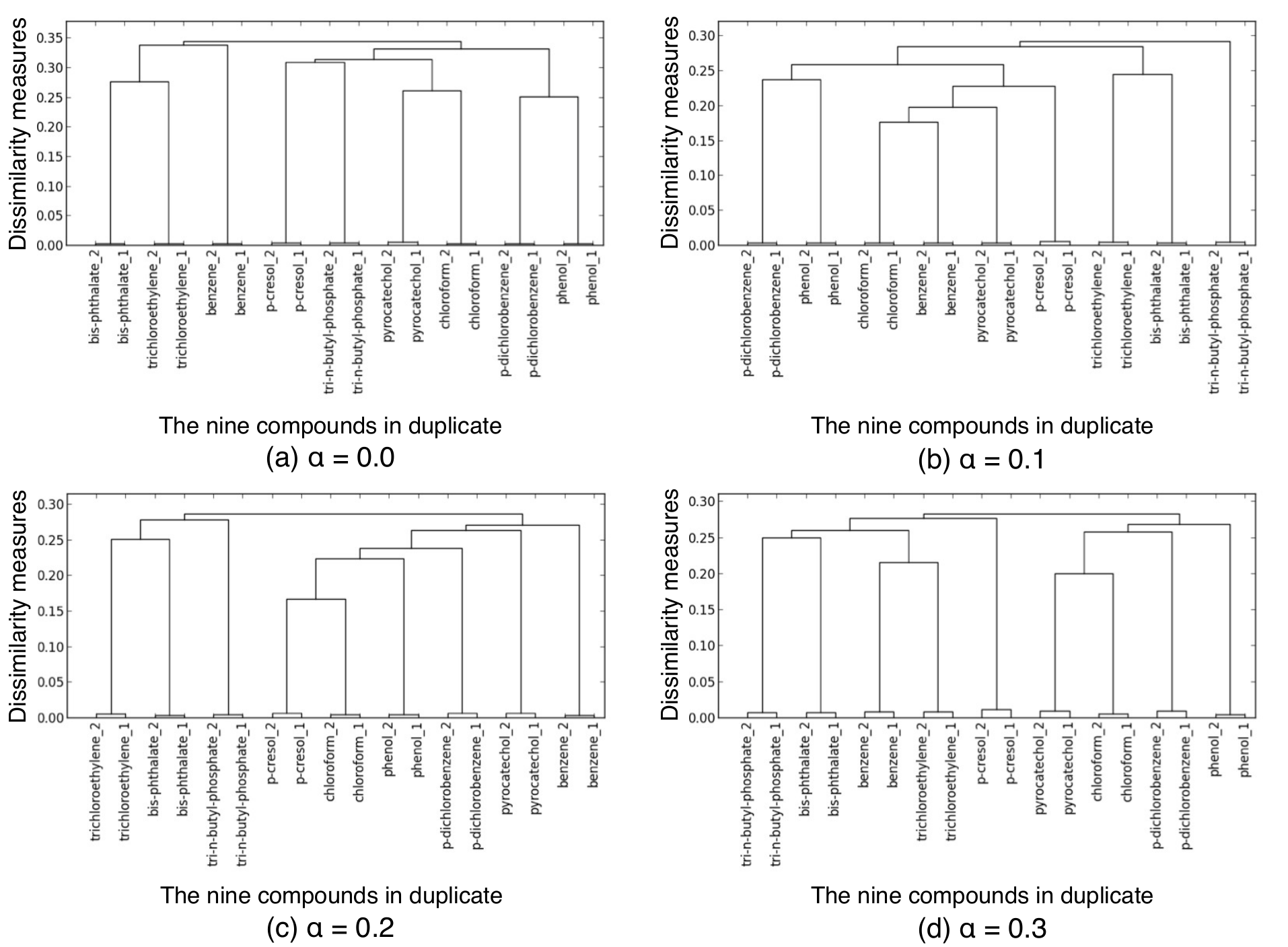}
\caption{Four dendrograms obtained by applying aggregative hierarchical clustering (the average linkage between the merged groups) to the data of the gene expression ratios for the nine compounds and $1,000$ extracted RNAs.
The upper-left (a), upper-right (b), lower-left (c) and lower-right (d) panels are the cases of $\alpha = 0.0, 0.1, 0.2$, and $0.3$, respectively.
In each panel, the $y$-axis marks the dissimilarity measures at which the clusters merge, and the $x$-axis the distribution of the nine compounds in duplicate.}
\label{fig:4}       
\end{figure}

\begin{figure}[htbp]
  \includegraphics[width=\textwidth]{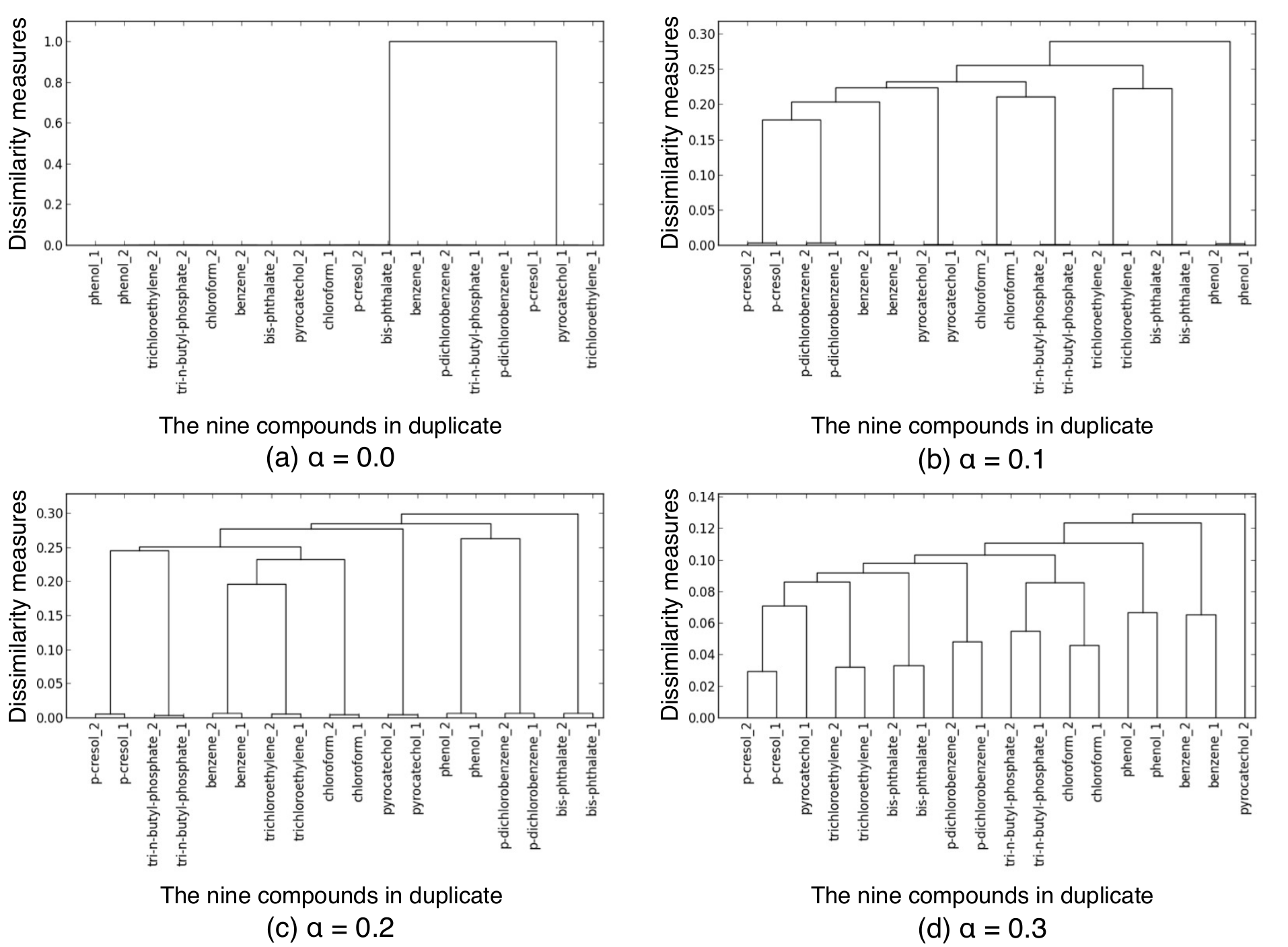}
\caption{Four dendrograms obtained by applying aggregative hierarchical clustering (the average linkage between the merged groups) to the data of the gene expression ratios for the nine compounds and $100$ extracted RNAs.
The upper-left (a), upper-right (b), lower-left (c) and lower-right (d) panels are the cases of $\alpha = 0.0, 0.1, 0.2$, and $0.3$, respectively.
In each panel, the $y$-axis marks the dissimilarity measures at which the clusters merge, and the $x$-axis the distribution of the nine compounds in duplicate.}
\label{fig:5}       
\end{figure}

We used the parameters $n = 3,000, 1,000, 100$, and $\alpha = 0.0, 0.1, 0.2, 0.3$, for the combinatorial optimization problem to extract RNAs.
Figures~\ref{fig:3}, \ref{fig:4}, and \ref{fig:5} show dendrograms obtained by agglomerative hierarchical clustering (the average linkage between the merged groups) applied to the data set with the nine compounds in duplicate, and $3,000$, $1,000$, and $100$ extracted RNAs, respectively.
Each figure has four panels. The upper-left (a), upper-right (b), lower-left (c), and lower-right (d) panes correspond to the cases of $\alpha = 0.0, 0.1, 0.2$, and $0.3$, respectively.
In each panel in these figures, the $y$-axis shows the dissimilarity measures where two clusters were merged, while the $x$-axis shows the distribution of compounds.
The meaning and function of the subscripts are as in Figures~\ref{fig:1} and \ref{fig:2}.

In Figures~\ref{fig:3}, \ref{fig:4}, and \ref{fig:5}, it is clear that the two samples are initially merged for all the compounds, and then different compounds are merged.
However, when $3,000$ RNAs were extracted (Figure~\ref{fig:3}), there are roughly $0.05$ dissimilarity measures between the two samples of each compound, although the maximum ranges are roughly $0.20$.
This result suggests that $3,000$ RNAs are not suitable for classifying compounds because the dissimilarity measures between the two samples of each compound are not small enough, compared to the dissimilarity measures between the compounds.
On the other hand, when $1,000$ or $100$ RNAs were extracted (Figures~\ref{fig:4} and \ref{fig:5}, respectively), and $\alpha = 0.1$ or $0.2$, the two samples are merged with sufficiently small dissimilarity measures, and we can clearly see the difference between compounds.
These results demonstrate the necessity of limiting the number of RNAs when classifying compounds to assess the effects of compounds on gene expression patterns.

\begin{figure}[htbp]
  \includegraphics[width=.75\textwidth]{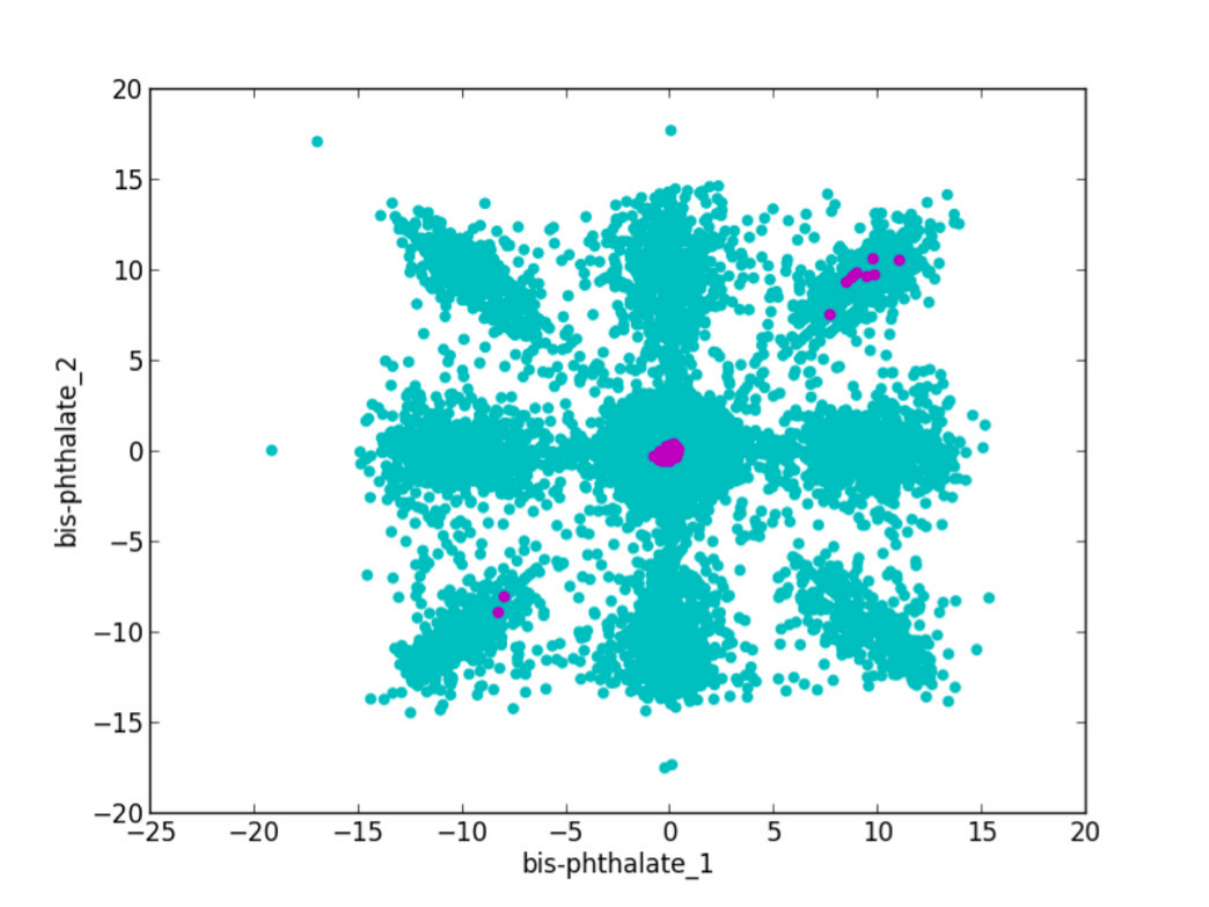}
\caption{Scatter plot of all the RNAs ($32,586$ RNAs) between the sample 1 and 2 in case of bis-phthalate.
The red plots indicate the $1,000$ extracted RNAs in case of $\alpha = 0.0$, and the blue plots are the rest RNAs.
The sizes of the extracted RNAs are almost zeros.}
\label{fig:6}       
\end{figure}

\begin{figure}[htbp]
  \includegraphics[width=.75\textwidth]{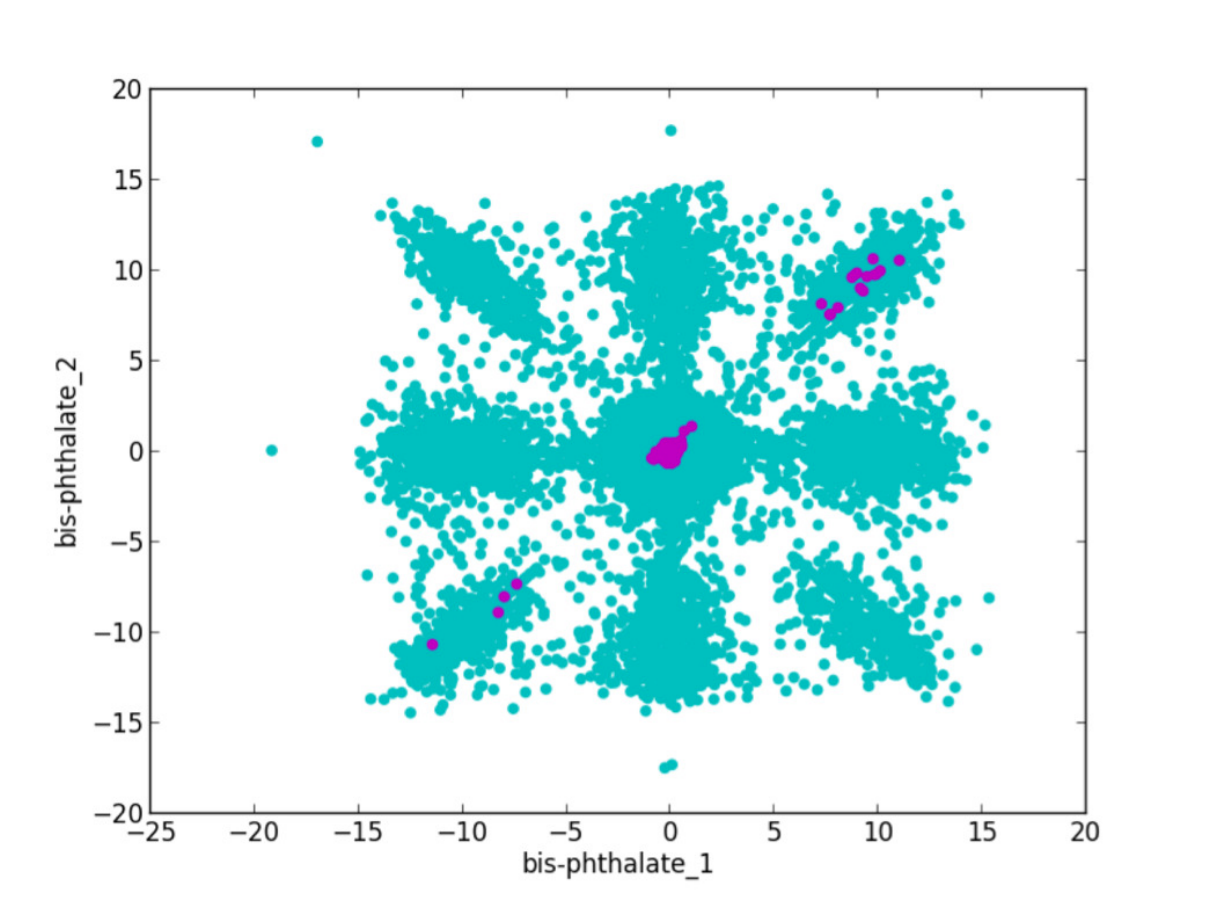}
\caption{Scatter plot of all the RNAs ($32,586$ RNAs) between the sample 1 and 2 in case of bis-phthalate.
The red plots indicate the $1,000$ extracted RNAs in case of $\alpha = 0.2$, and the blue plots are the rest RNAs.
The number of RNAs whose sizes are not zeros increase, compared to the case of $\alpha =0.0$.}
\label{fig:7}       
\end{figure}

Next, we compare the cases when $1,000$ and $100$ RNAs were extracted (Figures~\ref{fig:4} and \ref{fig:5}, respectively).
When $\alpha = 0.1$, neither case shows clear cluster structures;
however, when $\alpha = 0.2$, with $1,000$ extracted RNAs, the nine compounds are divided into two groups, one consisting of three compounds (bis-phthalate, trichloroethylene, and tri-n-butyl-phosphate) and the other of the remaining six compounds.
These results indicate that not only limiting the number of RNAs but also reflecting the sizes of RNAs is effective for revealing the difference between the compounds.
Figures~\ref{fig:6} ($\alpha = 0$) and \ref{fig:7} ($\alpha = 0.2$) show the scatter plots of all $32,586$ RNAs between Samples $1$ and $2$, in the case of bis-phthalate.
The red plots indicate the $1,000$ extracted RNAs, and the blue plots the remaining RNAs.
When $\alpha = 0.0$, it is clear that, for the most part, only RNAs with near-zero FPKMs are extracted;
whereas, when $\alpha = 0.2$, there is an increased number of RNAs of significant size.
Thus, using $n = 1,000$ and $\alpha = 0.2$ in the combinatorial optimization problem would be best for extracting RNAs in order to classify these compounds.

\section{Conclusion}
The present study has been applied agglomerative hierarchical clustering methods and a multiobjective combinatorial optimization problem to classify chemical compounds, based on \textit{in vitro} gene expression data.
This approach enables read-across based on \textit{in vitro} parameters, and the prediction of chemical toxicity.
However, even if we can properly classify compounds using \textit{in vitro} parameters, the results may not show the same similarity between compounds as in \textit{in vivo} experiments.
There is a gap in the respective similarities based on \textit{in vitro} parameters and \textit{in vivo} experiments.
Therefore, we must develop a means of extracting RNAs that reflects the similarity determined by \textit{in vivo} experiments, and this requires collaboration with toxicology experts.
This is a consideration for the future, but our approach would aid in such analysis.

\section*{acknowledgement}
  The authors would like to express appreciation to  Mr.\ Ryosuke Abe, Dr.\ Hiroshi Aoki, Dr.\ Hiroaki Sato, Dr.\ Masaki Torimura, and Dr.\ Masashi Gamo for thier fruitful discussions for the first version of manuscript.

\newpage

\section*{Figures}


\end{document}